\newcommand{\gama}{\chi}

\documentclass[oneside,a4paper,10pt]{amsart}
\usepackage{amsmath}               % AmSLaTeX
\usepackage{amscd}
\usepackage{amsthm}                % aggiunge nuovi ambienti tipo teorema
\usepackage{times}
\usepackage{graphicx}

\numberwithin{equation}{section}

\title[Non-spherical sources of static gravitational fields]%
{Non-spherical sources of static gravitational fields:
investigating the boundaries of the no-hair theorem}
\author[L. Herrera]{L. Herrera}
\address{Escuela de Fisica, Facultad de Ciencias, \hfill\break\indent Universidad
Central  de Venezuela,Caracas, Venezuela}
\email{laherrera@telcel.net.ve}
\author[G.\ Magli]{G. Magli}
\address{Dipartimento di Matematica,\hfill\break\indent Politecnico di
Milano, Italy} \email{magli@mate.polimi.it}
\author[D. Malafarina]{D. Malafarina}
\address{Dipartimento di Matematica,\hfill\break\indent Politecnico di
Milano, Italy} \email{malafarina@mate.polimi.it}

\begin{document}
\swapnumbers
\theoremstyle{plain}\newtheorem{teo}{Theorem}[section]
\theoremstyle{plain}\newtheorem{prop}[teo]{Proposition}
\theoremstyle{plain}\newtheorem{lem}[teo]{Lemma}
\theoremstyle{plain}\newtheorem{cor}[teo]{Corollary}
\theoremstyle{definition}\newtheorem{defin}[teo]{Definition}
\theoremstyle{remark}\newtheorem{rem}[teo]{Remark}
\theoremstyle{plain} \newtheorem{assum}[teo]{Assumption}
\theoremstyle{definition}\newtheorem{example}[teo]{Example}

\begin{abstract}

A new, globally regular model describing a static, non spherical
gravitating object in General Relativity is presented. The model
is composed by a vacuum Weyl--Levi-Civita special field - the so
called gamma metric - generated by a regular static distribution
of mass-energy. Standard requirements of physical reasonableness
such as, energy, matching and regularity conditions are satisfied.
The model is used as a toy in investigating various issues related
to the directional behavior of naked singularities in static
spacetimes and the blackhole (Schwarschild) limit.

\end{abstract}

\maketitle

\begin{section}{Introduction}\label{sec:intro}

Analytical modelling of astrophysical gravitating systems in
General Relativity is a poorly developed field, especially due to
the  difficulties occurring in the field equations when the -
obviously unrealistic - hypothesis of spherical symmetry is
abandoned. As a matter of fact we do not have {\it any} analytic
model which could resemble a realistic (i.e. rotating, axially
symmetric) star. This problem has an interesting theoretical
counterpart which is unsolved as well. It is the problem of
finding sources for the rotating blackhole solution under
conditions of physical reasonableness. Besides its obvious
intrinsic interest, finding this kind of solutions would allow
physicist to investigate those interesting phenomena which can be
expected to occur ``at the boundary'' between the singularity
theorems and the blackhole no-hair theorem. For instance, we do
not know anything about those processes which should produce the
progressive ``cuts of the hairs'' required to bring a non-vacuum,
non singular rotating and collapsing configuration to settle down
to a Kerr state before (naked) singularity formation. Of course
the same question arises in the problem of collapse of a
non-rotating source of a Schwarzschild black hole, which according
to the Israel theorem \cite{1} is the only static asymptotically
flat vacuum solution to the Einstein's equations with a regular
horizon. In order to delve deeper into these questions, we present
here an interior solution  which satisfies all physically relevant
requirements and  matches smoothly on the boundary surface to one
of the static axially symmetric vacuum solutions \cite{2} known as
the gamma metric  \cite{11}. This metric belongs to the family of
Weyl's solutions, and is continuously linked to the Schwarzschild
space-time through one of its parameters. The motivation for this
choice stems from the fact that the exterior gamma metric
corresponds to a solution of the Laplace equation (in cylindrical
coordinates) with the same singularity structure as the
Schwarzschild solution (a line segment). In this sense the gamma
metric appears as the ``natural'' generalization of Schwarzschild
space-time to the axisymmetric case. This new interior adds to the
few other known sources of the gamma metric \cite{7} and would
allow to study the behavior of very compact self-gravitating
systems, which according to the Israel theorem are very sensitive
to small fluctuations of spherical symmetry.

\end{section}

\begin{section}{Static, axially symmetric gravitational fields and the
gamma metric}\label{sec:nak}
The static, axially symmetric metric
can be written, in full generality, as
\begin{equation} \label{W}
 ds^{2}=-\gama^{2}dt^{2}+\alpha^{2}[(d{x}^{1})^{2}+(d{x}^{2})^{2}]+\beta^{2}
d\varphi^{2}
 \end{equation}
with $\alpha=\alpha(x^{1},x^{2}), \beta=\beta(x^{1},x^{2}),
 \gama=\gama(x^{1},x^{2})$.
The components of the Ricci tensor satisfy to
\begin{equation}\label{eq1}
 R^{t}_{t}+R^{\varphi}_{\varphi}=\frac{\triangle(\beta\gama)}{\alpha^{2}\beta\gama}
 \end{equation}
where $\triangle f= \frac{\partial
^{2}f}{\partial(x^{1})^{2}}+\frac{\partial
^{2}f}{\partial(x^{2})^{2}}$ is the {\it flat} Laplace operator in
the fictious two-dimensional space parameterized by cartesian
coordinates  $x^1$ and $x^2$.\\
Thus, if we are considering vacuum gravitational fields (so that
Ricci is zero) or if the matter source satisfies to
$T_{t}^{t}+T_{\varphi}^{\varphi}-T^\mu_\mu=0$ then
 one can make use of the so
called {\it Weyl gauge}. This consists in changing variables
$(x^{1},x^{2})\rightarrow(\varrho,z)$ where $\varrho :=\beta\gama$
is harmonic and $z$ is its harmonic conjugate. It follows that the
metric can be written as follows:
\begin{equation}\label{Weyl}
ds^{2}=-e^{2\lambda}dt^{2}+e^{2\nu-2\lambda}[(d\varrho)^{2}+(dz)^{2}]+\varrho^2e^{-2\lambda}d\varphi^{2}
 \end{equation}
 where
 \begin{equation}
 \alpha=e^{\nu-\lambda},\\
 \gama=e^{\lambda}.
 \end{equation}
 In the present section we shall consider only metrics satisfying the
 Weyl gauge (the physical content of this condition will be
 discussed in next section).
 The field equations can, in this case, be written as follows:
\begin{eqnarray}
\kappa
T_{t}^{t}&=&2\varrho\lambda_{,\varrho\varrho}+2\varrho\lambda_{,zz}+2\lambda
_{,\varrho}
-\varrho\lambda_{,\varrho}^{2}-\varrho\lambda_{,z}^{2}-\varrho\nu_{,\varrho\varrho}-\varrho\nu_{,zz}\\
\kappa
T_{\varrho}^{\varrho}&=&\varrho(\lambda_{,\varrho}^{2}-\lambda_{,z}^{2})-\nu
_{,\varrho}\\
\kappa
T_{z}^{z}&=&-\varrho(\lambda_{,\varrho}^{2}-\lambda_{,z}^{2})+\nu_{,\varrho}\\
\kappa T_{\varrho}^{z}&=&2\varrho\lambda_{,\varrho}\lambda_{,z}-\nu_{,z}\\
 \kappa
T_{\varphi}^{\varphi}&=&-\varrho(\lambda_{,\varrho}^{2}+\lambda_{,z}^{2}+\nu
_{,\varrho\varrho}+\nu_{,zz})
\end{eqnarray}
where we put
$$
\kappa:= 8\pi\varrho{e^{2\, \nu-2\,\lambda}}.
$$
It is worth recalling the structure of the vacuum solutions. For
vanishing energy-momentum tensor, we get
\begin{equation}\label{Laplace}
\Delta \lambda =0,
\end{equation}
\begin{equation}\label{nu}
\nu_{,\varrho}=\varrho(\lambda_{,\varrho}^{2}-\lambda_{,z}^{2}), \\
\nu_{,z}=2\varrho\lambda_{,\varrho}\lambda_{,z},
\end{equation}
\begin{equation}\label{nu2}
\Delta\nu+\lambda_{,\varrho}^{2}+\lambda_{,z}^{2}=0,
\end{equation}
where $\Delta$ is the flat Laplace operator on the fictitious
two-dimensional space parameterized by the cylindrical coordinates
$\varrho$ and $z$ (i.e. $\Delta f= f_{,\varrho \varrho} +f_{,zz}
+\frac{f_{,\varrho}}{\varrho}$). Once a solution of the Laplace
equation has been chosen, the second equation becomes an exact
differential and allows the calculation of $\nu$ by a quadrature:
\begin{equation}
\nu=\int_{\Gamma}\varrho[(\lambda_{,\varrho}^{2}-\lambda_{,z}^{2})d\varrho+2
\lambda_{,\varrho}\lambda_{,z}dz]+c
\end{equation}
with $\Gamma$ any open curve in the $(\varrho, z)$ plane, and it
can be verified that the third equation identically holds. There
is, therefore, a one to one correspondence between solutions of
the Laplace equation on flat two dimensional space and vacuum,
static, axially symmetric spacetimes. Asymptotic flatness requires
vanishing of $\lambda$ at space infinity, and the general solution
of Laplace equation is thus of the form \cite{HernandezMartin}:
\begin{equation} \label{Laplace2}
\lambda=\sum_{n=0}^{\infty}\frac{a_{n}}{R^{n+1}}P_{n}(\cos\psi)
\end{equation}
where
\begin{equation}
R=\sqrt{\varrho^{2}+z^{2}},\\ \cos\psi=\frac{z}{R}
\end{equation}
and $P_{n}(\cos\psi)$ are Legendre's polynomials. It then follows:
\begin{equation}
\nu=\sum_{n,k=0}^{\infty}\frac{(n+1)(k+1)}{n+k+2}\frac{a_{n}a_{k}}{R^{n+k+2}
}(P_{n+1}P_{k+1}-P_{n}P_{k}).
\end{equation}
The real constants $a_{n}$ are the so-called Weyl moments. The
Weyl moments are not the multipole moments measured by an observer
at space infinity \cite{ger}. For instance, the Schwarzsch\- ild
solution of mass $m$ has the following Weyl moments:
\begin{equation}
a_{2n}=-\frac{m^{2n+1}}{2n+1},\\ a_{2n+1}=0
\end{equation}
and thus it is not the monopole-Weyl solution. It is however easy
to relate the $a_n$ (Weyl) and $M_n$ (multipole) moments
\cite{fhp}; for instance it can be shown that
\begin{equation}
M_{0}= M = -a_{0},
\end{equation}
\begin{equation}
M_{2}= Q = a_{2}-\frac{1}{3}a_{0}^{3}.
\end{equation}
In what follows we shall concentrate on that particular metric
corresponding to the solution of the Laplace equation for a source
of constant density $\gamma$ uniformly distributed on a segment of
length $2m$ on  the $z$ axis centered at $z=0$ \cite{11}:
\begin{eqnarray}
\lambda&=&\frac{\gamma}{2}\ln\left(\frac{R_{+}+R_{-}-2m}{R_{+}+R_{-}+2m}\right)\\
\nu&=&\frac{\gamma^{2}}{2}\ln\left(\frac{(R_{+}+R_{-}-2m)(R_{+}+R_{-}+2m)}{4R_
{+}R_{-}}\right)
\end{eqnarray}
where $R_{+}=\sqrt{\rho^{2}+(z+m)^{2}}$,
$R_{-}=\sqrt{\rho^{2}+(z-m)^{2}}$. This metric is at best
visualized in the so-called Erez-Rosen coordinates
$(\rho,z)\rightarrow(r,\vartheta)$:
\begin{equation}
\rho^{2}=(r^{2}-2mr)\sin^{2}\vartheta, \\ z=(r-m)\cos\vartheta
\end{equation}
in which:
\begin{eqnarray}
\lambda&=&\frac{\gamma}{2}\ln\left(1-\frac{2m}{r}\right)\\
\nu&=&\frac{\gamma^{2}}{2}\ln\left(\frac{1-\frac{2m}{r}}{1-\frac{2m}{r}+\frac{
m^{2}}{r^{2}}\sin^{2}\vartheta}\right)
\end{eqnarray}
so that
\begin{equation}
ds^{2}=\Delta^{\gamma^{2}-\gamma-1}
\Sigma^{1-\gamma^{2}}dr^{2}+r^{2}\Delta^{\gamma^{2}-\gamma}\Sigma^{1-\gamma^
{2}}d\vartheta^{2}
+r^{2}\sin^{2}\vartheta\Delta^{1-\gamma}d\varphi^{2}-\Delta^{\gamma}dt^{2}
\end{equation}
where
\begin{equation}
\Delta=\left(1-\frac{2m}{r}\right),\\\Sigma=\left(1-\frac{2m}{r}+\frac{m^{2}
}{r^{2}}\sin^{2}\vartheta\right).
\end{equation}
This metric is usually called  {\it gamma metric }. It has many
interesting properties which we are going to discuss below.

\end{section}
\begin{section}{Physical properties of the gamma metric}\label{sec:phs}

The gamma metric coincides with the Schwarzschild metric of mass
$m$ when $\gamma=1$. One of the advantages of the Erez-Rosen frame
is that in this limit the gamma reduces to Schwarzschild in
Schwarzschild coordinates. The mass and the quadrupole moment of
the gamma metric are given by \cite{HP}:
\begin{equation}
 M=\gamma m,
 \end{equation}
\begin{equation}
 Q=(\gamma^{2}-1)\frac{\gamma}{3}m^{3}.
\end{equation}
In the $\gamma=0$ limit the metric becomes flat, whilst the Curzon
solution (i.e. the Weyl monopole solution) is obtained in the
limit $\gamma\rightarrow\infty$, $m\rightarrow0$ with $\gamma m
= const$ \cite{Pap}.\\
As we have seen the gamma metric is originated - formally - by the
solution of the Laplace equation for a line bar of constant
density. However this a rather formal identification, and we are
interested here in properties of the gamma metric as a vacuum
field observed at space infinity. First of all, one can study the
surface of revolution analogue to
Schwarzschild spheres \cite{Newman}.\\
To an $r=r_b$ surface in Schwarzschild pertains an area
\begin{equation}
A_{S}=\int^{\pi}_{0}\int^{2\pi}_{0}r^{2}_{b}\sin\vartheta
d\vartheta d\varphi = 4\pi r^{2}_{b} \ .
\end{equation}
Calculating {\it the same} area for $r=r_b$ surfaces in the gamma
case, we can classify such surfaces  as oblate or prolate with
respect to the Schwarzshild sphere. Obviously, this property of
the revolution surfaces is also the expected ``shape'' of static
axially symmetric sources of the metric.\\
 The 2-dimensional metric on
the slices $r=r_b$, $t=const$ is
\begin{equation}
d\Omega^{2}=r^{2}_{b}\Delta_{b}^{\gamma^{2}-\gamma}\Sigma_{b}^{1-\gamma^{2}}
d\vartheta^{2}
+r^{2}_{b}\sin^{2}\vartheta\Delta_{b}^{1-\gamma}d\varphi^{2}
\end{equation}
where a subscript denotes evaluation at $r=r_b$. One has, to
lowest order in $\epsilon=\gamma-1$,
\begin{equation}
det\bar{g} =
r_{b}^{4}\sin^{2}\vartheta\Delta_{b}^{(1-\gamma)^{2}}\Sigma_{b}^{1-\gamma^{2}}
= r_{b}^{2}\sin\vartheta(1-\epsilon \ln\Sigma_{b}).
\end{equation}
From which follows:
\begin{equation}
A_{\gamma}=A_{S}\left(1-\frac{1}{2}\epsilon\int_{0}^{\pi}\sin\vartheta
\ln\Sigma_{b} d\vartheta\right).
\end{equation}
Since $\ln\Sigma_{b}<0$ we get $A_{\gamma}>A_{S}$ for $\epsilon>0$
($\gamma>1$); in such cases the slices are oblate (respectively
prolate for $\gamma<1$). If the metric has to be interpreted has
the vacuum field generated by a symmetric object then the
physically interesting case is obviously that of oblate sources
(e.g. galaxies). Stability has, for instance, to be expected only
for such objects. This result is in agreement with the physical
shape since the expansion of the newtonian potential is given by
\cite{ger}:
\begin{eqnarray}\nonumber
 g_{00}&=&-1+2\left[\frac{M_{0}}{r}+\frac{Q}{r^{3}}P_{2}(\cos\vartheta)+o\left
(\frac{1}{r^{3}}\right)\right]
 \\ &=&-1+2\left[\frac{\gamma m}{r}+\frac{1}{r^{3}}\frac{\gamma
m^{3}}{3}(\gamma^{2}-1)
 P_{2}(\cos\vartheta)+o\left(\frac{1}{r^{3}}\right)\right]
\end{eqnarray}
so that the newtonian potential of the source is that of a disk if
$\gamma>1$.

\end{section}

\begin{section}{The nature of the singularity of the gamma metric}\label{sec:n}

In the present section we investigate on the nature of the
singularity of the gamma metric, extending and completing previous
results by Virbhadra \cite{vib}. The Kretschmann scalar
$\textsl{K}=R_{abcd}R^{abcd}$ is
\begin{equation}
\textsl{K}(r,\theta)=\frac{16m^{2}\gamma^{2}N(r, \vartheta
)}{r^{10}\Delta^{2\gamma^{2}-2\gamma+2}\Sigma^{3-2\gamma^{2}}}
\end{equation}
where
\begin{eqnarray}
\nonumber N(r, \vartheta
)&:=&m^{2}\sin^{2}\vartheta[3m\gamma(\gamma^{2}+1)(m-r)+\gamma^{2}(4m^{2}-6mr+
3r^{2})+\\
 &&+m^{2}(\gamma^{4}+1)]+3r(m\gamma+m-r)^{2}(r-2m).
\end{eqnarray}
For $\gamma=1$ we get the Schwarzschild result
$\textsl{K}=\frac{48m^{2}}{r^{6}}$, which is obviously regular at
$r=2m$. In what follows we consider only the case $\gamma\neq 1$.
The $r=2m$ surface is then a true singularity of the spacetime and
the no-hair theorem assures that it will be a naked singularity in
the sense that it is not covered
by an event horizon.\\
The diverging behavior of $\textsl{K}$ in $r=2m$ obviously depends
on $\vartheta$. For $\vartheta=0$ one has $\Sigma=\Delta$ and
therefore:
\begin{equation}
\textsl{K(r,0)}=\frac{48m^{2}\gamma^{2}(m\gamma+m-r)^{2}}{r^{8}\Delta^{4-2\gamma}},
\end{equation}
this function diverges at $r=2m$ only if $\gamma<2$.\\
If $\vartheta=\vartheta_0\neq 0$
$K(r,\vartheta_{0})\approx\Delta^{-\gamma^{2}+\gamma-1}$. Since
$\gamma^{2}-\gamma+1>0$ $\forall\gamma\neq0$, the Kretschmann
scalar diverges in $r=2m$ if the singularity is approached on any
plane different from the polar one, for any value of $\gamma$. It
is therefore interesting to investigate the visibility of such
singularity in
dependence of the direction in which it is approached. \\
 Consider a null ``radial'' ($\vartheta=const$) geodesic. A
 faraway observer can receive a signal from the $r=2m$ surface if
\begin{equation}
\lim_{{\varepsilon\rightarrow0}}\int^{{R}}_{{2m+\varepsilon}}dt=
\lim_{{\varepsilon\rightarrow0}}\int^{{R}}_{{2m+\varepsilon}}
\Delta^{(\frac{\gamma^{2}-2\gamma-1}{2})}\Sigma^{\frac{1-\gamma^{2}}{2}}dr
<\infty.
\end{equation}
On any plane different from the polar one ($\vartheta\neq0$) there
exists a positive constant $C$ such that
\begin{equation}
\lim_{{\varepsilon\rightarrow0}}\int^{{R}}_{{2m+\varepsilon}}
\Delta^{(\frac{\gamma^{2}-2\gamma-1}{2})}\Sigma^{\frac{1-\gamma^{2}}{2}}dr<
\lim_{{\varepsilon\rightarrow0}}C\int^{{R}}_{{2m+\varepsilon}}
\Delta^{(\frac{\gamma^{2}-2\gamma-1}{2})}dr<\infty
\end{equation}
since the integral on the right converges for any
$\gamma\neq1$ (if $\gamma=1$ divergence of the integral obviously signals
the presence of the event horizon).\\
If however $\vartheta=0$:
\begin{equation}
\lim_{{\varepsilon\rightarrow0}}\int^{{R}}_{{2m+\varepsilon}}
\Delta^{(\frac{\gamma^{2}-2\gamma-1}{2})}\Sigma^{\frac{1-\gamma^{2}}{2}}dr=
\lim_{{\varepsilon\rightarrow0}}\int^{{R}}_{{2m+\varepsilon}}
\Delta^{-\gamma}dr
\end{equation}
and the integral converges if and only if $\gamma<1$. The
situation is summarized in the table 1. We conclude that the
singularity is naked independently on the direction of observation
if $\gamma<1$ while it exhibits directional nakedness for
$\gamma>1$, being not visible, in this case, on the polar plane.
\begin{table} [hhh!]
\begin{large}
  \centering
  \begin{tabular}{c|c|c}
      & $\gamma<1$ & $\gamma>1$ \\ \hline
    $\vartheta=0$ & visible & not visible \\ \hline
    $\vartheta\neq 0$ & visible & visible
  \end{tabular}
   \medskip \caption{\textit{Directional behavior of the singularity of the gamma metric}}
  \end{large}
\end{table}

\vspace{-0.5cm}

\section{Sources of the gamma metric}

To construct physically viable sources of the gamma in order to
investigate their behavior ``at the boundary of the no-hair
theorem'' one needs to chose a matter model and then integrate the
coupled Einstein+matter field equations. However, this proved by
now to be impossible in the non-spherical case (also in the case
of the Schwarzschild field few exact interior solutions are
known). We thus need a simplifying assumption which however should
restrain as little as possible the physical significance of the
results. There is, in fact, the famous argument known as ``Synge
trick'' which proposes to ``guess''  the metric from general
considerations and then to evaluate the matter content from
Einstein equations; however, usually such ``Synge solutions'' have
nothing to do with physics. The idea we follow here traces back to
work by Hernandez \cite{Hernandez} and it is based on deformations
of physically valid spherically symmetric solutions. We interpret
here these deformations as actual changes in the shape of the
objects, i.e. we use $\epsilon$ as the deformation parameter.\\
As far as we are aware the unique known examples of interior
solution for the gamma are that given by Stewart et al. \cite{7}.
One of this solution is  based on the Schwarzschild constant
density interior solution as a ``seed'', but the resulting density
is not strictly decreasing from the center at higher orders and
therefore the solution cannot be stable with respect to radial
perturbations. The second one is based on the Adler interior
solution.\\
 In order to construct a new interior solution we take
here a different point of view. Rather than searching for a
solution generated by a ``seed'' to lowest order, we consider a
``variation of the mass'' of the gamma metric, namely, that family
of interior solutions which is obtained by (a priori) any
inhomogeneous distribution of mass which generates the vacuum
gamma field at a fixed boundary. It is obvious, that not all the
mass distributions will be physically viable, and indeed we shall
see that the space of allowed functions is severely restricted by
physical conditions.\\
It is well known that an interior solution generating the
Schwarzschild field can be obtained from the vacuum metric putting
$m=\mu(r)$. The Einstein tensor then becomes:
 \begin{eqnarray}
  G^{0}_{0}&=&2\frac{\mu'(r)}{r^{2}} \\
  G^{1}_{1}&=&2\frac{\mu'(r)}{r^{2}} \\
  G^{2}_{2}&=&\frac{\mu''(r)}{r}\\
  G^{3}_{3}&=&\frac{\mu''(r)}{r}
\end{eqnarray}
 So that the matching, regularity and energy conditions
\begin{eqnarray}
  \mu(r_{b})&=&m\\
  \lim_{r\rightarrow 0} \frac{\mu(r)}{r^{2}}&=&0\\
2\frac{\mu'}{r^{2}}&\geq&0\\
2\frac{\mu'}{r^{2}}-\frac{\mu''}{r}&\geq&0
\end{eqnarray}
 can easily be satisfied, for instance, by a matter distribution of the form
 \begin{equation}\label{schint}
  \mu(r)=\frac{4}{3}\pi
  E_{0}r^{3}\left(1-\frac{3}{4}\frac{r}{r_{b}}\right).
\end{equation}
 Note that the energy condition $G^{0}_{0}-G^{1}_{1}\geq0$ holds
identically for all
 $r$.\\
 We now apply the same reasoning to the gamma. Considering
 $\gamma=1+\epsilon$ to first order in $\epsilon$ the vacuum
 metric becomes:
 \begin{eqnarray}
  g_{00}&=&-\Delta-\epsilon\Delta\ln\Delta \\
  g_{11}&=&\Delta^{-1}+\epsilon\Delta^{-1}\left(\ln\Delta-2\ln\Sigma\right) \\
  g_{22}&=&r^{2}+\epsilon r^{2}\left(\ln\Delta-2\ln\Sigma\right)\\
  g_{33}&=&r^{2}\sin^{2}\vartheta-\epsilon
  r^{2}\sin^{2}\vartheta\ln\Delta
\end{eqnarray}
 So that taking the Schwarzschild interior solution as the order zero
interior solution is
 possible to build an interior of the form
 \begin{eqnarray}\label{GammaInt}
\tilde{g}_{00}&=&-\left(1-\frac{2\mu(r)}{r}\right)-\epsilon\left(1-\frac{2\mu(r)
}{r}\right)F(r) \\
  \tilde{g}_{11}&=&\left(1-\frac{2\mu(r)}{r}\right)^{-1}
+\epsilon\left(1-\frac{2\mu(r)}{r}\right)^{-1}\left(F(r)+G(r,\vartheta)\right)
\\
  \tilde{g}_{22}&=&r^{2}+\epsilon r^{2}\left(F(r)+G(r,
  \vartheta)\right)\\ \label{GammaInt2}
  \tilde{g}_{33}&=&r^{2}\sin^{2}\vartheta-\epsilon r^{2}\sin^{2}\vartheta F(r)
\end{eqnarray}
 Matching conditions require
 \begin{eqnarray}
  \mu(r_{b})&=&m \\
  \mu'(r_{b})&=&0 \\
  F(r_{b})&=& \ln\left(1-\frac{2m}{r_{b}}\right) \\
  F'(r_{b})&=& \frac{2m}{r_{b}^{2}}\left(1-\frac{2m}{r_{b}}\right)^{-1} \\
  G(r_{b}, \vartheta)&=&
-2\ln\left(1-\frac{2m}{r_{b}}+\frac{m^{2}}{r_{b}^{2}}\sin^{2}\vartheta\right)=A(\vartheta) \\
  G'(r_{b}, \vartheta)&=& -\frac{4m}{r_{b}^{2}}\frac{1-\frac{m}{r_{b}}\sin^{2}\vartheta}
  {1-\frac{2m}{r_{b}}+\frac{m^{2}}{r_{b}^{2}}\sin^{2}\vartheta}=B(\vartheta)
\end{eqnarray}
 while regularity conditions for $r\rightarrow 0$ require
 \begin{eqnarray}\label{match}
  \frac{\mu(r)}{r^{2}}& \longrightarrow & 0 \\
  \frac{G(r, \vartheta)}{r^{2}} & \longrightarrow & 0
\end{eqnarray}
 Now the energy conditions have to be imposed. It turns out that a suitable
 solution which satisfies the weak energy conditions
 can be found only for $\epsilon>0$, i.e. for oblate sources.\\
 The first energy condition, namely $G^{0}_{0}\geq 0$, is satisfied
 automatically for all $r \neq r_{b}$ if the corresponding condition for
 Schwarzschild is satisfied. In $r=r_{b}$ the order zero condition vanishes
 and so $[G^{0}_{0}]_{r=r_{b}}\geq 0$ must be imposed. From this
 follows:
 \begin{equation}\label{G''}
  G''(r_{b},
  \vartheta)\leq\left(1-\frac{2m}{r_{b}}\right)^{-1}\left(\frac{4m}{r_{b}^{3}}
  -\frac{1}{r_{b}^{2}}G_{\vartheta\vartheta}(r_{b},\vartheta)-\frac{1}{r_{b}}
  \left(1-\frac{m}{r_{b}}\right)G'(r_{b},\vartheta)\right).
 \end{equation}
 In the same way conditions $G^{0}_{0}-G^{2}_{2}\geq 0$ and
$G^{0}_{0}-G^{3}_{3}\geq
 0$ are satisfied automatically for all $r \neq 0$ and so only
 $[G^{0}_{0}-G^{2}_{2}]_{r=0}\geq 0$ and $[G^{0}_{0}-G^{3}_{3}]_{r=0}\geq
0$ must
 be imposed.
The first is however satisfied by means of the regularity
condition for $G(r, \vartheta)$ while the second leads to
\begin{equation}
  \lim_{r\rightarrow 0}\left(2\frac{F'(r)}{r}+F''(r)\right)= c\geq0.
\end{equation}
Finally the condition that must be satisfied for every $r$ and
$\vartheta$ is
\begin{eqnarray}\nonumber
G^{0}_{0}-G^{1}_{1}&=&\left(1-\frac{2\mu(r)}{r}\right)\left(\frac{2}{r}F'(r)-\frac{1}{2}G''(r,\vartheta)\right)+
  \\
  &&-\frac{1}{2r^{2}}\left(G_{\vartheta\vartheta}(r,\vartheta)
  +\frac{\cos\vartheta}{\sin\vartheta}G_{\vartheta}(r,\vartheta)\right)\geq0.
\end{eqnarray}
 It is easy to check that the latter is satisfied in $r=0$ and
$r=r_{b}$ if the previous conditions are satisfied and so, by
continuity it must be satisfied in a neighborhood of those two
points. Taking an appropriate function $G(r,\vartheta)$ so that
$G''(r, \vartheta)$ and $G_{\vartheta\vartheta}(r,\vartheta)
  +\frac{\cos\vartheta}{\sin\vartheta}G_{\vartheta}(r,\vartheta)$
are small compared to $\frac{F'(r)}{r}$ for all $\vartheta$
satisfies the condition everywhere. Explicit examples of choices
of $F(r)$ and $G(r,\vartheta)$ satisfying the conditions are given
in the appendix. However we show in the figures the behavior of
the energy density and the critical energy condition for one of
such examples describing a compact object with $m=1.4M_{\bigodot}$
and $r_{b}=10Km$.
\begin{center}
\begin{figure}[h]
  \hfill
  \flushleft
  \begin{minipage}[t]{.45\textwidth}
\includegraphics{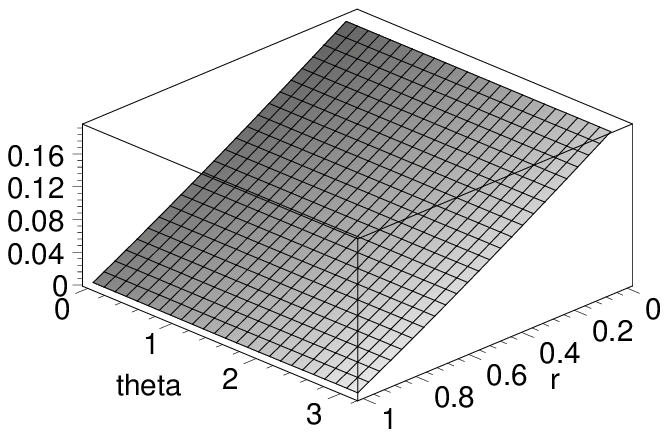}
      \center{(a)}
  \end{minipage}
  \hfill
  \begin{minipage}[t]{.45\textwidth}
      \includegraphics{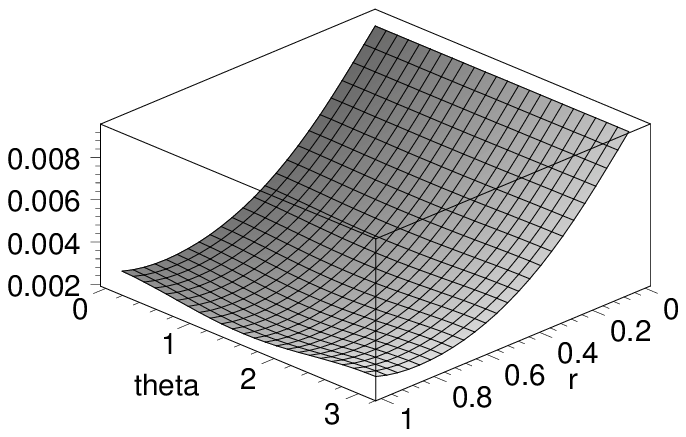}
      \center{(b)}
  \end{minipage}
  \hfill
  \caption{\textit{In figure are shown (a) the condition for the energy density $T^{0}_{0}\geq0$
  and (b)
  the condition $T^{0}_{0}-T^{1}_{1}\geq0$ (in the adimensional unit obtained multiplying by $r_{b}^{2}$)
  for a fixed value of $\epsilon=\frac{1}{100}$ in dependence of $\vartheta$ and the adimensional
  radial variable $\frac{r}{r_{b}}$}.}
\end{figure}
  \end{center}

\end{section}

\begin{section}{Discussion and conclusions}\label{sec:n}

General Relativity is in principle the ``final theory'' for the
description of macroscopic gravitating objects. It is therefore
odd enough that after nearly a century from its invention we do
not have even one sound model of a star within the theory. In fact
no solution of stationary axially symmetric Einstein field
equation in matter is known, which is able to represent a finite
non singular rotating object. Even in the simpler case of static,
but not spherical, gravitating objects very little is known. Of
course this case is the non rotating limit of the previous one,
since it is only in the case of the perfect fluid that the static
object must be spherical, and rotating stars are not to be
expected to be made out of perfect fluids. For instance it is very
well known that anisotropy plays a relevant role
\cite{HerreraSantos} in the structure of compact objects such as
white dwarfs and neutron
stars.\\
So motivated we investigated here a simple static but not
spherically symmetric model of a compact object, showing the
existence of physically valid solutions of Einstein field
equations describing it.\\
This model should allow for studying the dynamic behavior of the
compact object close to the surface $r_{b}=2m$. This may be done,
either by considering the dynamic equation just after the system
leaves the equilibrium \cite{Herrera} or by  calculating the
active gravitational (Tolman) mass as the system leaves the
equilibrium \cite{Herrerabis}. In either case the bifurcation
between the exactly spherically symmetric case and the gamma
solution, should appear even for very small values of
$\gamma$, if $\frac{m}{r_{b}}$ is sufficiently close to $\frac{1}{2}$.\\
However, observe that specific constraints appear in our model on
the value of $\frac{m}{r_{b}}$. Indeed, the central  radial
pressure is given by
\begin{equation} \nonumber
P(0)=E_{0}+\epsilon\left(-E_{0}\ln\left(1-\frac{2m}{r_{b}}\right)+\frac{F'(r
_{b})}{r_{b}}\left
(\frac{1}{2}E_{0}r_{b}^{2}-\frac{1}{8\pi}\right)
 +\frac{C}{r_{b}^{2}}\left(\frac{1}{6}E_{0}r_{b}^{2}-\frac{1}{8\pi}\right)\right)
\end{equation}
which is consistent with the small departure from spherical
symmetry approximation if
\begin{equation}\nonumber
\epsilon\left(-E_{0}\ln\left(1-\frac{2m}{r_{b}}\right) +\frac{F'(r
_{b})}{r_{b}}\left(\frac{1}{2}E_{0}r_{b}^{2}-\frac{1}{8\pi}\right)
 +\frac{C}{r_{b}^{2}}\left(\frac{1}{6}E_{0}r_{b}^{2}-\frac{1}{8\pi}\right)\right) \ll
 E_{0}.
\end{equation}
Since both $\ln(1-\frac{2m}{r_{b}})$ and $\frac{F'(r
_{b})}{r_{b}}=\frac{2m}{r_{b}^{3}}\left(1-\frac{2m}{r_{b}}\right)^{-1}$
diverge as $\frac{m}{r_{b}}\rightarrow\frac{1}{2}$, the pressure
blows up in that limit. We must therefore consider an upper limit
in the surface gravitational potential for which the model is
acceptable. The same reasoning can be applied to the ``arbitrary''
positive constant $C$. In fact once the scale given by $\epsilon$
is fixed the value of $C$ must not exceed a
limit value for which the approximation holds.\\
Given the fact that $\frac{m}{r_{b}}=\frac{1}{3}\pi
E_{0}r_{b}^{2}$ maybe it is better to say that once we fix the
value of $\frac{m}{r_{b}}$ (and so $E_{0}$) the possible values of
$\epsilon$ become fixed as well. So that for
$\frac{m}{r_{b}}\rightarrow\frac{1}{2}$ we must have
$\epsilon\rightarrow0$, and the more we depart from spherical
symmetry the less the value allowed for $\frac{m}{r_{b}}$
becomes.\\
Finally we would like to mention that work aimed to extend these
results to slow rotation scenario is in progress.

\end{section}

\appendix
\section{Example}
 The Einstein tensor for the interior metric \ref{GammaInt}-\ref{GammaInt2} results:
 \begin{eqnarray} \nonumber
  G^{0}_{0}&=&2\frac{\mu'(r)}{r^{2}}
+\epsilon\left[-2\frac{\mu'(r)}{r^{2}}(F+G)-\frac{G_{\vartheta\vartheta}}{2r^{
2}}
  -\frac{1}{2}\left(1-\frac{2\mu(r)}{r}\right)G''+\right.
  \\
&&\left.+\frac{1}{2r}\left(\mu'(r)-\frac{\mu(r)}{r}\right)G'+\frac{1}{2r}\left(1
-\frac{2\mu(r)}{r}\right)(2F'-G')\right] \\ \nonumber
  G^{1}_{1}&=&2\frac{\mu'(r)}{r^{2}}
+\epsilon\left[-2\frac{\mu'(r)}{r^{2}}(F+G)+\frac{\cos\vartheta}{\sin\vartheta
}\frac{G_{\vartheta}}{2r^{2}}+\right.
  \\
  &&\left.+\frac{1}{2r}\left(\mu'(r)-\frac{\mu(r)}{r}\right)G'-\frac{1}{2r}\left(1-\frac{2\mu(r)}{r}\right)(2F'+G')\right]\\
  \nonumber
  G^{2}_{2}&=&\frac{\mu''(r)}{r}
+\epsilon\left[-\frac{\mu''(r)}{r}(F+G)-\frac{\cos\vartheta}{\sin\vartheta}\frac{G_{\vartheta}}{2r^{2}}+\right.
  \\
&&\left.-\frac{1}{2r}\left(\mu'(r)-\frac{\mu(r)}{r}\right)G'+\frac{1}{2r}\left(1
-\frac{2\mu(r)}{r}\right)(2F'+G')\right]
\\ \nonumber
  G^{3}_{3}&=&\frac{\mu''(r)}{r}
  +\epsilon\left[-\frac{\mu''(r)}{r}(F+G)-\frac{G_{\vartheta\vartheta}}{2r^{2}}
  -\frac{1}{2}\left(1-\frac{2\mu(r)}{r}\right)(2F''+G'')+\right.
  \\
&&\left.+\frac{1}{2r}\left(\mu'(r)-\frac{\mu(r)}{r}\right)(4F'+G')-\frac{1}{2r}\left(1-\frac{2\mu(r)}{r}\right)(2F'+G')\right]
\end{eqnarray}
A simple choice of $F(r)$ satisfying all conditions is
\begin{equation}
  F(r)=\frac{1}{2}\frac{F'(r_{b})}{r_{b}}r^{2}+\frac{1}{2}C\frac{r^{2}}{r_{b}^{2}}\left(1-\frac{2}{3}\frac{r}{r_{b}}\right)+D
\end{equation}
with $C$ an arbitrary
positive constant and $D$ given by the condition for $F(r)$ at the boundary.\\
 Furthermore considering $G(r,\vartheta)$ of the form
\begin{eqnarray}
  G(r, \vartheta)&=&g(r)A(\vartheta)+h(r)B(\vartheta)
\end{eqnarray}
 with $g(r)\geq0$ so that $g(r_{b})=1$ and $g'(r_{b})=0$ and $h(r)\leq0$ so that
$h(r_{b})=0$ and $h'(r_{b})=1$ satisfies the
 matching conditions. The regularity requires:
\begin{equation}
\lim_{r\rightarrow 0}\frac{g(r)}{r^{2}}=\lim_{r\rightarrow
0}\frac{h(r)}{r^{2}}=0.
\end{equation}
The condition $G^{0}_{0}-G^{1}_{1}\geq0$ is then satisfied if
$g(r)$, $g''(r)$, $h(r)$ and $h''(r)$ are small enough with
respect to $F(r)$. Which is always possible due to the degree of
freedom given by the choice of the positive constant $C$.
 The condition $[G^{0}_{0}]_{r=r_{b}}\geq 0$ finally implies
\begin{equation}
  G''(r_{b},\vartheta)\leq
\left(1-\frac{2m}{r_{b}}\right)^{-1}\left(\frac{4m}{r_{b}^{3}}-\frac{A_{\vartheta\vartheta}(\vartheta)}{r_{b}^{2}}
  -\left(1-\frac{m}{r_{b}}\right)\frac{B(\vartheta)}{r_{b}}\right)
\end{equation}
which is easily satisfied.\\ If we consider an object with
$m=1.4M_{\bigodot}$ and $r_{b}=10Km$ (i.e. a neutron star) in
geometrized units we get $\frac{m}{r_{b}}=\frac{1}{5}$ and
$E_{0}=\frac{3}{5\pi r_{b}^{2}}$ (which corresponds to
$2.6\cdot10^{15}\frac{g}{cm^{3}}$) then we have a wide variety of
choices for the functions $F(r)$ and $G(r, \vartheta)$ and the
central pressure doesn't diverge for every small value of
$\epsilon$ so that the reasonable choice $\epsilon=\frac{1}{100}$
leads to a central pressure $\frac{1}{1000}$ times greater than
the corresponding pressure for the spherically symmetric case.

\end{document}